\documentclass[a4paper,useAMS,usenatbib,12pt,fleqn]{mn2e}

\usepackage[T1]{fontenc}
\usepackage{ae,aecompl}

\usepackage{epsf,graphicx,lscape,subfigure,longtable,txfonts,wrapfig,float,
}
\usepackage{rotating}
\usepackage{amssymb}
\usepackage[light]{}
\usepackage{multicol}
\usepackage{footnote}
\usepackage{apalike}
\usepackage{caption}

\def\kms{\ifmmode{\rm km\,s}^{-1}\, \else km\,s$^{-1}$\,\fi}
\def\mujy{$\mathrm{\muup Jy\,}$}

\def\mjybm{mJy\,beam$^{-1}$}
\def\ltsim{\ifmmode\stackrel{<}{_{\sim}}\else$\stackrel{<}{_{\sim}}$\fi}
\def\gtsim{\ifmmode\stackrel{>}{_{\sim}}\else$\stackrel{>}{_{\sim}}$\fi}
\def\S4195{41.95+575}
\def\S4331{43.31+592}

\def\solmasyr{$\rm{M_\odot yr^{-1}}$}

\title[ALMA observations of Wd1-9]{ALMA observations of the supergiant B[e] star Wd1-9}
\author[D. M. Fenech  et al.]{D. M. Fenech$^1$, J. S. Clark$^2$, R. K. Prinja$^1$, J. C. Morford$^1$, S. Dougherty$^3$, R. Blomme$^4$\\
$^{1}$Dept. of Physics {\&} Astronomy, University College London, Gower Street, London WC1E 6BT \\
$^{2}$Dept. Physical Sciences, The Open University, Walton Hall, Milton Keynes, MK7 6AA  \\
$^{3}$Dominion Radio Astrophysical Observatory, National Research Council Canada, PO Box 248, Penticton, B.C.  V2A 6J9 \\
$^4$Royal Observatory of Belgium, Ringlaan 3, 1180 Brussel, Belgium \\
}

\date{Accepted XXX. Received YYY; in original form ZZZ}

\pubyear{2016}

\begin{document}
\label{firstpage}
\pagerange{\pageref{firstpage}--\pageref{lastpage}}
\maketitle

\begin{abstract}
{ Mass-loss in massive stars plays a critical role in their evolution, although the precise mechanism(s) responsible - radiatively driven winds, impulsive ejection and/or binary interaction - remain uncertain. In this paper we present ALMA line and continuum observations of the supergiant B[e] star Wd1-9, a massive post-Main Sequence object located within the starburst cluster Westerlund 1. We find it to be one of the brightest stellar point sources in the sky at millimetre wavelengths, with (serendipitously identified) emission in the H41$\alpha$ radio recombination line. We attribute these properties to a low velocity ($\sim100$\,\kms) ionised wind, with an extreme mass-loss rate  $\gtrsim$6.4$\times10^{-5}(d/5kpc)^{1.5}$\,\solmasyr. External to this is an extended aspherical ejection nebula  indicative of a prior phase of significant mass-loss. Taken together, the millimetre properties of Wd1-9  show a remarkable similarity to those of the highly luminous stellar source MWC349A.
We conclude that these objects are interacting binaries evolving away from the main sequence and undergoing  rapid case-A mass transfer. As such they - and by extension the wider class of supergiant B[e]  stars - may provide a unique window into the physics of a process that shapes the life-cycle of $\sim70$\% of massive stars found in binary systems.}

\end{abstract}


\begin{figure*}
\begin{center}
\includegraphics[width=15cm]{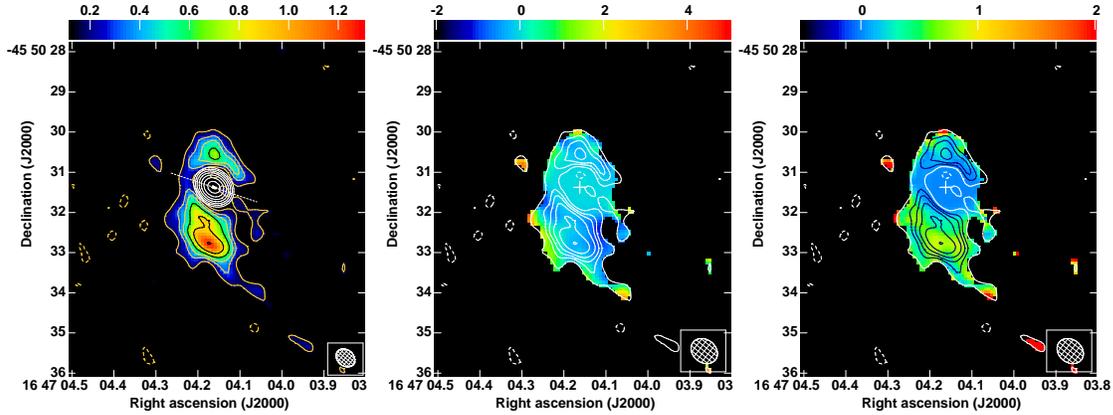} 
\caption{Continuum images at 100GHz of Wd1-9. Left: Robust-0 weighted image of the extended continuum emission shown in colour-scale and gold contours. The contours are plotted at 155\,\mujy (3$\sigma$) $\times$ -1,1,2,2.8,4,5.7,8,11.3,16 from a R0 weighted image (beam size: 0.51$\,\times$ 0.40$''$). The compact source emission is also displayed in white contours plotted at 1.4\,\mjybm$\times$ 0.1,0.14,0.2,0.28,0.4,0.57,0.8. The dashed line shows the orientation of the line emission centroids (see Section \ref{linetext} for further details). The middle and right images show the spectral index and spectral index error images of Wd1-9 respectively from a naturally weighted MFS clean (beam size: 0.58$\,\times$ 0.46$''$). The white contours are the same as those plotted in gold in the left-hand image showing the extended emission. The cross shows the peak position of the compact source.}
\label{wd19}
\end{center}
\end{figure*}

\section{Introduction}

 The evolution of massive OB stars is dominated by mass-loss throughout their life-cycle, with the nature of their death - core-collapse supernova versus prompt collapse - and ultimate fate (neutron star or black hole) dependent on how much matter they are able to shed. Historically, line-driven winds have been thought to mediate this mass-loss. However, there is a significant uncertainty in our understanding of this phenomenon; depending on whether such winds are smooth or clumped, the resultant  mass-loss rates of massive stars are in question at the order-of-magnitude level \citep[e.g.][]{fullerton06,puls06}. 
Given this discord, other physical mechanisms have been invoked to explain how the outer mantle of O stars may be stripped to yield H-depleted Wolf-Rayet (WR) stars, with impulsive mass-loss driven by instabilities in the luminous blue variable (LBV) or yellow hypergiant/red supergiant phase a prime candidate \citep[e.g.][]{humphrey79,abr14}. However, recent analysis has shown that fully 70\% of massive stars are in binaries that will interact at some point in their life-cycle \citep{demink14}, also resulting in significant mass-loss from both the primary to the secondary as well as from the binary system as a whole. 

 In order to place observational constraints on these disparate processes we obtained millimetre continuum observations of the massive Galactic starburst cluster Westerlund 1 \citep[Wd1;][]{clark05} with the Atacama Large Millimetre/Submillimeter Array (ALMA). Observations at such wavelengths are particularly sensitive to wind-clumping via thermal Bremsstrahlung emission \citep[e.g.][]{blomme03}, while ejection nebula formed via impulsive mass-loss events or binary interaction may also be identified and resolved at millimetre-radio wavelengths.
With an age of $\sim5$\,Myr, Wd1 is the ideal laboratory to undertake such a study since it contains a uniquely rich population of massive, evolved single and binary stars at a distance that enables the detection of several tens of objects (Fenech et al. in prep.). In this study we focus on the millimetre properties of the supergiant (sg)B[e] star Wd1-9 (R.A. 16$^{{h}}$47$^{m}$4.162$^{s}$ Dec. -45$^{o}$50$'$31.373$''$ ). 
SgB[e] stars are characterised by an aspherical circumstellar environment, comprising a high velocity polar wind and a warm, dusty and gaseous equatorial disc/torus \citep{zickgraf85}, leading to  a pronounced near-IR continuum excess and rich emission line  spectrum. The origin of the disc is uncertain, but recently interest has grown in binarity as a physical driver of the phenomenon \citep{pods06,clark13a}; if this is the case then the circumstellar environment may encode information on the mass-loss history of the binary and critically the physics driving it. 

Wd1-9 is the only sgB[e] star known to reside in a star cluster and hence one of the handful of Galactic objects for which a post-MS (main sequence) evolutionary state may be unambiguously determined \citep{clark13b}. Interest in it was reignited by the discovery that it is one of the most luminous stellar radio sources known \citep{clark98,dough10}. A synthesis of the  multi-wavelength and epoch dataset compiled as a result of this finding  was presented in \cite{clark13b},  which revealed an emission spectrum arising from the circumstellar environment that apparently entirely veils the nature of the central source, although the IR excess clearly indicates the presence of a hot dusty torus. Additionally, the presence of emission lines from high-excitation species (e.g. [S\,{\sc iv}], [O\,{\sc iv}]) and an unexpected X-ray luminosity \citep{clark08} provides persuasive evidence for binarity.

\section{Observations and image processing}

Observations of Westerlund 1 were performed with ALMA on 30th June - 1st July 2015 (Project code: 2013.1.00897.S). The observations were made at a central frequency of 100\,GHz with a total bandwidth of 7.5\,GHz over four spectral windows. Each window contains 128 channels with a frequency width of 15.625\,kHz. In total 27 pointings were used to cover Westerlund 1, however the results presented here utilise data solely from the pointing centred on Westerlund 1 \#9 (Wd1-9). The array utilised 42 antennas with baselines ranging from 40 to 1500\,m and a total on-source integration time per pointing of 242.3\,secs ($\sim$4\,mins). The data were calibrated using the standard ALMA pipeline procedures in Common Astronomy Software Applications (CASA) and included application of apriori calibration information as well flagging of erroneous data. Observations of J1617-5848 were used to perform the phase and bandpass calibration and observations of Titan and Pallas were used to amplitude calibrate the data assuming flux densities of 228.96 and 82.01\,mJy respectively at 91.495\,GHz. Self-calibration was then carried out using the bright ($\sim$150\,mJy) compact source emission from Wd1-9.\\
\indent In addition to the observed continuum emission we serendipitously detected broad line emission coincident with the compact emission in Wd1-9 at a frequency of 92.060\,GHz which we identify as the H41$\alpha$ radio recombination line (RRL).  In order to analyse this RRL emission, following calibration, we analysed a single spectral window (91.42-93.42\,GHz) containing the line emission following subtraction of the continuum contribution. \\
\indent Imaging of both the continuum and line emission was performed using the CLEAN algorithm in CASA. Multiple images of the continuum emission were produced using Cotton-Schwab cleaning with both natural weighting (beam size: 0.58$\times$ 0.46$''$) and Briggs Robust-0 (R0, beam size: 0.51$\,\times$ 0.40$''$) weighting. Imaging was also performed using the multi-frequency synthesis (MFS) capability for both of these weighting schemes to produce spectral index and associated error maps. This performs frequency-dependent modelling during deconvolution to spatially map the spectral index and associated errors using the full 7.5\,GHz bandwidth. All spectral indices and errors presented are taken directly from the spectral index and error images shown in Fig. \ref{wd19} \citep[for further details see][]{rau11}. \\
\indent In order to thoroughly study the low surface brightness extended emission of Wd1-9 and mitigate the high dynamic range requirements, images of the extended emission were made following subtraction of the strong compact source emission from the visibility data. In all cases, images of the continuum emission were produced excluding the channels containing the line emission. An image cube was also produced in the local standard of rest (LSR) frame using the continuum-subtracted data with natural weighting and a velocity resolution of 50.90\,\kms.

\section{Wd1-9 morphology}

\subsection{Continuum emission}

 The ALMA continuum imaging clearly shows two distinct components associated with Wd1-9; a bright compact source surrounded by lower brightness extended emission. Fig. \ref{wd19} shows a R0-weighted image of the extended emission with overlayed contours of the compact source emission from an equivalent image. 
A 2-D gaussian fit to the compact source component gives a deconvolved size of 158.3$\pm$1.0 $\times$ 121.9$\pm$1.4\,mas and a position angle (P.A.) of 0.7$\pm$1.2\,$\mathrm{^{o}}$\footnote{P.A. is defined E through N or anti-clockwise} with an integrated flux density of 153$\pm$8\,mJy. 

 Previous ATCA observations of Westerlund 1 at 8.6\,GHz published by \cite{dough10} also showed both compact and extended emission components of Wd1-9. They determined a spectral index for the compact source of +0.68$\pm$0.07. A simple extrapolation from the observed radio flux density \citep[24.9\,mJy at 8.6\,GHz;][]{dough10} predicts a 100\,GHz flux density of $132.1^{+24.7}_{-20.8}$\,mJy consistent with the 153\,mJy observed. However, the compact source emission in the millimetre regime, whilst thermal, appears to have a much flatter spectral index, with a value of 0.000$\pm$0.015 
at the peak of the compact emission, as can be seen from the spectral index map in Fig. \ref{wd19}. This would potentially indicate an opacity transition between the radio, where the wind was partially optically thick \citep[as noted by][]{dough10}, and the millimetre where it appears to be perhaps more optically thin.  


 The resolved low surface brightness emission of Wd1-9 shows an almost N-S orientation with a P.A. $\sim$0\,$\mathrm{^{o}}$ (in good agreement with the deconvolved P.A. of the compact source) extending $\sim$1.4\,arcsec to the north and $\sim$1.9\,arcsec to the south, with the southerly lobe showing additional substructure. At an assumed distance of 5\,kpc to Wd1-9 \footnote{Throughout the paper we adopt a distance of 1.4\,kpc to MWC349A  and 5kpc to Wd1-9.} this implies a total extension of $\sim$0.08\,pc. This N-S structural asymmetry could be a result of a system inclination (see Sect. \ref{disc}), which \cite{clark13b} calculate to be $\sim$50$\mathrm{^{o}}$. It is also possible that this asymmetry could be the result of interaction with the cluster wind. However, given the core of the cluster lies to the south of Wd1-9 and the southern component is the more prominent, it is more likely a function of increased ionising flux, as has previously been suggested by \cite{dough10} to be the cause of the cometary nebulae around several other stars within Wd1. The morphology of the extended emission clearly deviates from the spherical assumption made by \cite{dough10} when analysing the extended radio emission, though a direct comparison is difficult because of the differing capabilities of the ALMA and ATCA instruments.
The spectral index image in Fig. \ref{wd19} shows the extended emission also appears to be thermal. The N and S peak components within the maps have an index of -0.2$\pm$0.2 and -0.3$\pm$0.8 respectively. This as for the radio spectral index \citep[0.16$\pm$0.07; ][]{dough10} indicates optically thin emission.

\begin{figure}
\begin{center}
\includegraphics[height=59mm]{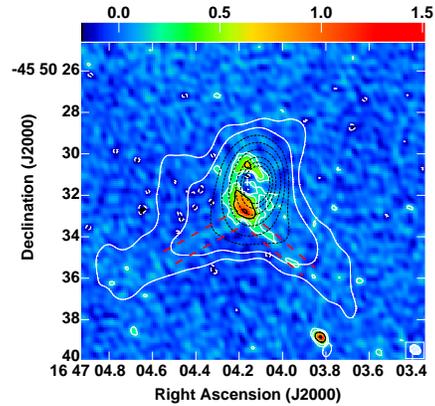}
\caption{\small{Background image is Robust-0 weighting clean image of Wd1-9 from the ALMA observations (following subtraction of the compact source from the data). Contours are as listed in Fig. \ref{wd19}. The overlying contours are from the 1998 ATCA observations at 8.6\,GHz (see Dougherty et al. 2010). The red dashed lines indicate the low brightness filaments.}}
\label{wd19rad}
\end{center}
\end{figure}

 It is also interesting to note the presence of two narrow filaments extending to the SE and SW of the source. The SE component is clearly visible in the image in Fig. \ref{wd19} with a $\>3\mathrm{\sigma}$ detection. The SW component is somewhat fainter, appearing at $\sim 2\mathrm{\sigma}$ (see Fig. \ref{wd19rad}). Similar structure is seen in the 8.6\,GHz ATCA observations \citep[][]{dough10} suggesting these structures are real, however more sensitive follow-up observations are required to confirm this.

\subsection{H41$\mathrm{\alpha}$ RRL emission}\label{linetext}

\begin{figure*}
\begin{center}
\includegraphics[width=8.0cm,height=5.7cm]{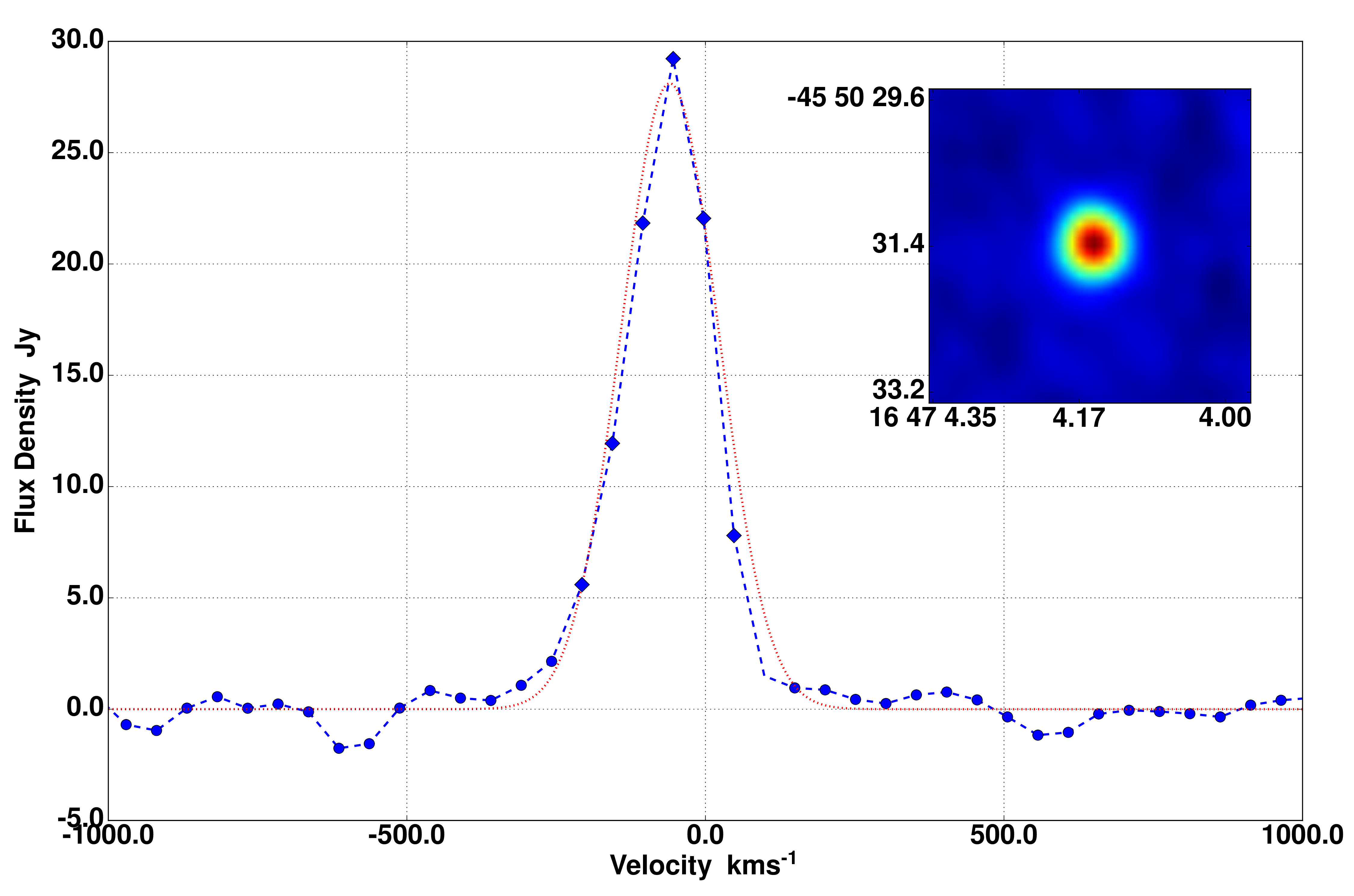} 
\includegraphics[width=9.6cm,height=5.8cm]{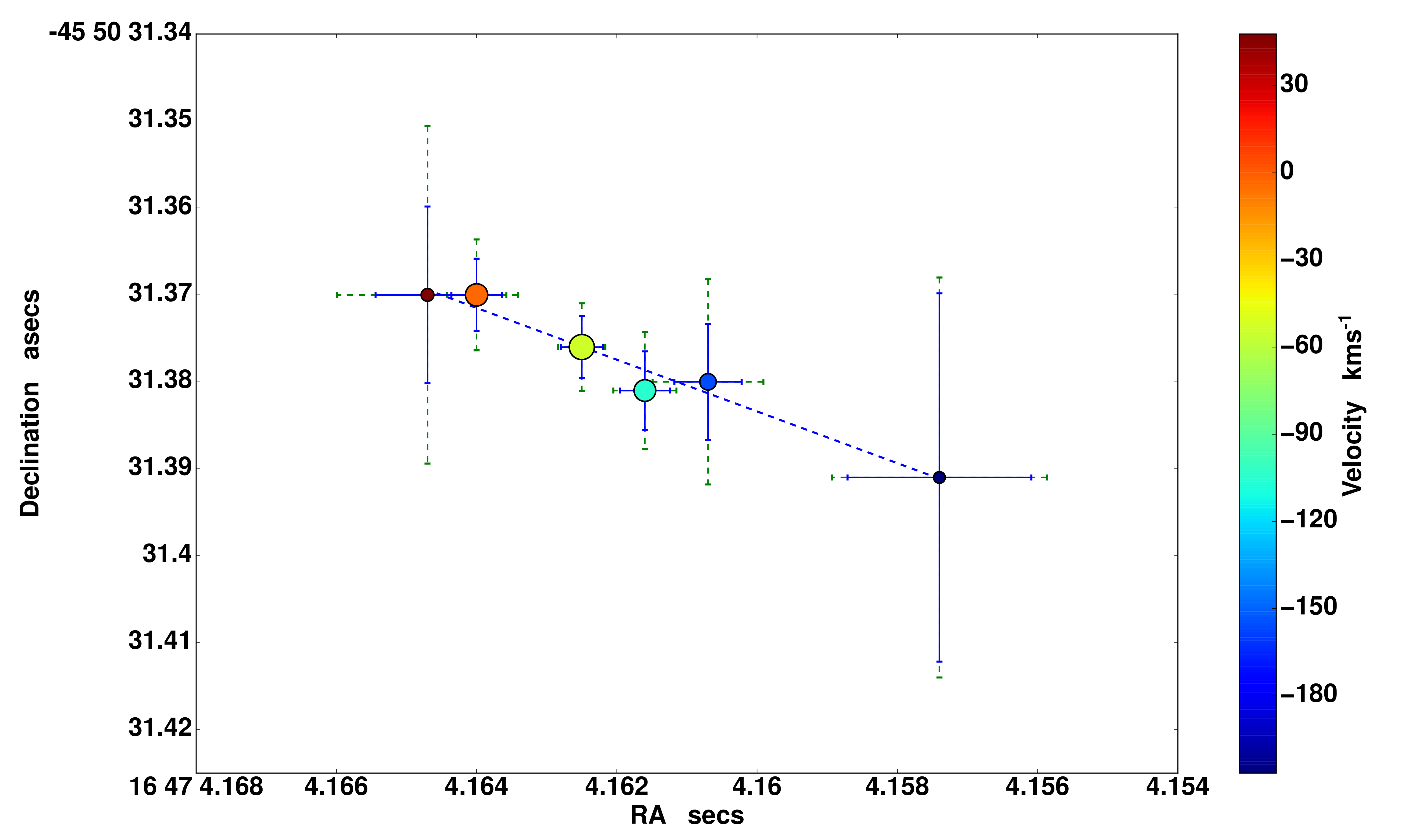} 
\caption{Left: H41$\alpha$ RRL profile. Blue dots and diamonds depict the data. The Gaussian fit to the profile is shown as the red dashed line. The inset colour-scale image shows the unresolved RRL emission in the peak channel (beam size: 0.69\,$\times$0.63$''$). Right: Fitted centroid positions of the line emission for each channel plotted with a diamond in the left-hand profile plot. Gaussian fitted positions with errors are shown in blue. The colour-scale represents the frequency and the point sizes show the relative intensity of the emission. Also shown in green dashed lines are formal position errors (see Section \ref{linetext} for further details). The blue dashed line shows a linear fit to the centroid positions and is also displayed in Fig. \ref{wd19}.}
\label{lineplots}
\end{center}
\end{figure*}

The RRL emission appears spatially as an unresolved point source coincident with the compact component of the continuum emission. The RRL profile is plotted in Fig. \ref{lineplots} and shows a broad single peak with a gaussian-fitted FWHM of 163$\pm$5\,\kms. The peak velocity -59$\pm$2\,\kms is consistent with the cluster velocity for Westerlund 1 \citep{clark14}. 
Also plotted in Fig. \ref{lineplots} are the peak positions of the line emission as observed in each channel of the image cube. The positions were determined using a 2-D gaussian fit and the errors shown are the gaussian-fitting calculated error and the formal position error as outlined in \cite{weintroub08}. 
This diagram suggests that the position of the centroids vary linearly with a P.A. 71$\mathrm{^{o}}$ i.e. approximately perpendicular to the orientation of the extended emission, as illustrated by the dashed line in Fig. \ref{wd19}. As can be seen from both the RRL profile and the channel centroids plotted in Fig. \ref{lineplots}, there appears to be a total velocity gradient of $\sim$200\,\kms associated with the RRL emission. The $\sim$80\,mas shift seen in RA, equates to $\sim$400\,AU (at 5\,kpc) and is directly comparable to the scale of the dusty torus inferred by \cite{clark13b}.

\section{Discussion}\label{disc}

At mm wavelengths Wd1-9 is one of the most extreme  stellar sources known. In terms of intrinsic luminosity, the only comparators are the LBV $\eta$ Carinae \citep{abr14} and the B[e] star MWC349 \citep{white85}, where the emission arises, respectively,  from a compact ejection nebula and a dense wind. Moreover, as with Wd1-9 both these stars are amongst a  handful of stellar objects which also exhibit mm/radio recombination line emission \citep[e.g.][]{abr14,martin89}. Mindful of  the putative association between LBVs and sgB[e] stars \citep[e.g.][]{clark13a} this immediately raises the question of which source (if either) does Wd1-9 resemble.

  The spatial variability of the gaussian H41$\alpha$ emission line is consistent with an origin in either a bipolar outflow or a rotating disc. In MWC349A this line is similarly single peaked and is thought to arise in a disc wind outflow \citep[e.g.][]{martin89}, whereas the higher transition (e.g. H30$\alpha$) line profiles appear double-peaked and are believed to originate in a rotating circumstellar disc \citep[e.g.][]{weintroub08,baez13}. Given the marked multiwavelength (optical-radio)  similarities between Wd1-9 and MWC349A  \citep[e.g.][]{clark13b} we assume the same for the H41$\alpha$ RRL here, although we caution that further observations will be required to unambiguously distinguish between both options.

If correct, the width of the line profile suggests an outflow velocity of $\sim100$\,\kms for Wd1-9 and, in conjunction with both the continuum mm and radio flux densities and the emission model from \cite{dough10}, we calculate a current mass-loss rate of \.{M}$\sim$\,6.4${\times}10^{-5}(d/5kpc)^ {1.5}$\,\solmasyr \,\footnote{Stellar parameters are  those taken from \cite{dough10} and spherical geometry is assumed.}. Both parameters are atypical  for the late O/early B supergiants found within Wd1 \citep{neg10} and which might be expected  to characterise the properties of the polar wind of a sgB[e] star \citep[following][]{zickgraf85}.  Indeed, such a slow and dense wind is only encountered amongst rather extreme examples of LBVs \citep{clark14}.  However, this does not necessarily favour comparison of Wd1-9 to $\eta$ Carinae as \cite{abr14} attribute  the continuum flux of $\eta$ Carinae to a shell ejection event with \.{M}$\sim10^{-2.5}-10^{-1.5}$\,\solmasyr;  over two orders of magnitude greater than we infer for Wd1-9. 

However, the wind parameters for Wd1-9 are broadly comparable to those of MWC349A  which, utilising the fluxes and $v_{\infty}$ $\sim$\,50\,\kms from \cite{tafoya04}, gives \.{M}$\sim$\,1.8${\times}10^{-5}(d/1.4kpc)^{1.5}$\,\solmasyr assuming spherical symmetry and a distance of 1.4\,kpc \citep{rygl12}.  Following \cite{tafoya04}, accounting for the non-spherical wind in MWC349A leads to a reduction in mass loss rate by a factor of two; a similar adjustment would also be required for Wd1-9 if future observations reveal it too shares a similar wind morphology, leading to final mass loss rates of $\sim$\,0.9${\times}10^{-5}(d/1.4kpc)^{1.5}$\,\solmasyr\, and $\sim$\,3.2${\times}10^{-5}(d/5kpc)^{1.5}$\,\solmasyr for MWC349A and Wd1-9 respectively.
Unlike the standard sgB[e] paradigm where the high velocity polar outflow is a standard line-driven stellar wind, in MWC349A it is attributed to a disc wind from the ionised  surface of the predominantly neutral (un-ionised) circumstellar torus \citep{planesas92}. This is apparent in the bipolar `hourglass' radio morphology \citep[e.g.][]{tafoya04} which is bisected by the opaque dusty torus. While the presence of a dusty torus is inferred for Wd1-9 via modelling of its IR  continuum \citep[inner/outer radii $\sim$60\,AU/800\,AU respectively with $\sim$50$\mathrm{^{o}}$ inclination;][]{clark13b}, the greater distance to it precludes directly resolving either this or the putative disc wind with {\em current} observations (both structures reside within our observed compact component). \\
\indent Next we turn to the extended emission in Wd1-9 and recent mid-IR observations of MWC349A which suggest a remarkably similar nebula is present in its system \citep{gvar12}. In the case of MWC349A the mid-IR nebula has a more developed `hourglass' morphology, with the waist of the emission orientated in an E-W direction. The lobes of this nebula in MWC349 are outlined by approximately straight filaments extending from each end of the waist of the emission, suggesting the possible faint narrow filaments associated with Wd1-9 may represent a similar morphology. It is interesting to note however that the full development of a parsec-scale nebula (as seen for MWC349) around Wd1-9 could be restricted because of the high pressure of the ambient cluster medium within Wd1. High spatial resolution near-IR observations of the more compact neutral torus in MWC349A reveal that it also shares an E-W orientation \citep{danchi01}, with the bipolar radio emission that is attributed to a disc wind lying either side of this feature, with a resultant N-S orientation. If our hypothesis that MWC349A and Wd1-9 are similar systems is correct, we would attribute the spatially extended N-S band of emission in Wd1-9 (Fig. \ref{wd19}) to a similar toroidal structure seen (approximately) edge-on. Furthermore if the compact point source embedded within it also represents a bipolar wind one would expect it to demonstrate an E-W orientation - a hypothesis apparently supported via our spectro-astrometry of the H41$\alpha$ line.\\ 
\indent Intriguingly the massive interacting binary RY Scuti also supports a comparable nebular morphology, which has been attributed to impulsive mass-loss events driven and shaped by binary interaction \citep[][]{gehrz95,smith11}.  As described in Sect. 1, the X-ray luminosity and spectrum  of Wd1-9 clearly argues for binarity; comparison to other cluster members with comparable properties and orbital solutions \citep[e.g. WR A \& B, Wd1-13]{bon07, ritchie10} would suggest a short orbital period ($<15$\,days) comparable to that of RY Scuti ($\sim$11.125\,d).  We therefore propose that the outflow, torii and nebulae around Wd1-9 are also the result of binary interaction as has been suggested for MWC349 \citep{gvar12}.

\section{Concluding remarks}

We have presented new ALMA 3mm observations of Wd1-9 that reveal it to be exceptionally luminous, suggesting  an extreme mass-loss rate in comparison to other cluster members. We identify both compact and extended thermal components and unexpected line emission in the H41$\alpha$ transition associated with the former.  Analysis of the line centroids implies a velocity gradient of $\sim200$\,\kms perpendicular to the bright N-S component of the extended nebula. \cite{clark13b} have previously highlighted the close similarity between Wd1-9 and the B[e] star MWC349A at optical and IR wavelengths; these observations extend such a comparison to the mm and radio regime. Both systems are thought to host compact circumstellar torii which, in the case of MWC349A drives a massive, slow wind perpendicular to its surface. By analogy to MWC349A we suggest that the H41$\alpha$ line emission associated with Wd1-9 likewise arises in such an outflow. {\em If} this hypothesis is correct, the terminal velocity and mass-loss rates of the winds in both systems are comparable to within a factor of $\sim4$. Moreover, if Wd1-9 and MWC349A host similar circumstellar environments we would expect that future optimised observations will reveal higher transitions (e.g. H29-31$\alpha$) that are double peaked \citep[cf.][]{weintroub08,baez13} and delineate the dusty torus (running N-S) perpendicular to our putative outflow. 

Assuming Wd1-9 and MWC349A are similar systems we may draw several far reaching conclusions. Firstly MWC349A would be unambiguously a post-MS object, resolving a decades-old debate \citep[cf.][ and refs. therein]{gvar12}. Secondly both systems would form prime laboratories for the study of disc-wind launching, of considerable interest to the study of massive protostars as well as post-MS evolution. Similarly, both would provide a window into the poorly understood physics of massive interacting binaries. The mass-loss rates inferred for both Wd1-9 and MWC349A are sufficiently large that they cannot be attributed to binary interaction on the nuclear ($\sim10^6$yr) timescale since this would yield a total mass lost to the systems that was directly comparable to the initial masses inferred for  the primaries. Instead we suppose that we are observing rapid case A mass transfer in both systems occuring on the much shorter thermal timescale (few $10^4$yr). In such a scenario a percentage of this mass is accreted by the binary, with the remainder lost to the binary system \citep{petrovic05}; both MWC349 and Wd1-9 suggest that the latter  process is mediated via a low-terminal-velocity disc wind.

Extending this conclusion, we note that while formally classifiable as sgB[e] stars following the criteria of e.g. \cite{zickgraf85}, both Wd1-9 and MWC349A appear to differ from other examples in the Magellanic Clouds. In both cases this is due to (i) the origin of the polar wind (disc-driven rather than a classical stellar wind) and (ii) the high temperature and compact nature of the dusty toroids. We speculate that these differences result from the stage of binary interaction in which we observe these systems, with extensive mass transfer and mass-loss occurring within Wd1-9 and MWC349A on a thermal timescale, while archetypal sgB[e] stars such as R126 \citep{kastner10} with colder more extended discs may represent post mass-transfer systems. 

Finally, as  noted before, the high mass loss rates and low wind velocities that characterise both Wd1-9 and MWC349A are similar to the outflow properties of LBVs -  both inferred directly and from analysis of ejection nebulae \citep{clark14}. Given the LBV-like variability observed in some sgB[e] binaries \citep{clark13a}, one might speculate that the extreme mass loss rates of some  LBV-like systems are also driven by binarity (cf. RY Scuti).

\section*{Acknowledgements}

This paper makes use of the following ALMA data: ADS/JAO.ALMA\#2013.1.00897.S. ALMA is a partnership of ESO (representing its member states), NSF (USA) and NINS (Japan), together with NRC (Canada), NSC and ASIAA (Taiwan), and KASI (Republic of Korea), in cooperation with the Republic of Chile. The Joint ALMA Observatory is operated by ESO, AUI/NRAO and NAOJ. D. Fenech and J. Morford wish to acknowledge funding from a STFC consolidated grant (ST/M001334/1) and a STFC studentship respectively.

\label{lastpage}

\begin{thebibliography}{99}


\bibitem[\protect\citeauthoryear{Abraham {et al.}} {2014}]{abr14} Abraham Z., Falceta-Gon\c{c}alves D. \& Beaklini P., 2014, ApJ, 791, 95 
\bibitem[\protect\citeauthoryear{ B\'{a}ez-Rubio {et al.}} {2013}]{baez13} B\'{a}ez-Rubio A., Mart\'{i}n-Pintado J., Thum C. \& Planesas P., 2013, A\&A, 553, A45
\bibitem[\protect\citeauthoryear{Bonanos} {2007}]{bon07}Bonanos A., 2007, AJ, 133, 2696

\bibitem[\protect\citeauthoryear{Blomme {et al.}} {2003}]{blomme03} Blomme R., van de Steene G., Prinja R., Runacres M., Clark J., 2003, A\&A, 408, 715
\bibitem[\protect\citeauthoryear{Clark {et al.}} {1998}]{clark98} Clark J., Fender R., Waters L. B., Dougherty S., Koornneef J., Steele I. \& van Blokland A., 1998, MNRAS, 299, L43
\bibitem[\protect\citeauthoryear{Clark {et al.}} {2005}]{clark05} Clark J., Negueruela I., Crowther P. \& Goodwin S., 2005, A\&A, 434, 949

\bibitem[\protect\citeauthoryear{Clark {et al.}} {2008}]{clark08} Clark J., Muno M., Negueruela I., Dougherty S., Crowther P., Goodwin S. \& de Grijs R., 2008, A\&A, 477, 147

\bibitem[\protect\citeauthoryear{Clark {et al.}} {2013a}]{clark13a} Clark J., Bartlett E., Coe M., Dorda R., Haberl F., Lamb J., Negueruela I., Udalski A., 2013a A\&A, 506, A10

\bibitem[\protect\citeauthoryear{Clark {et al.}} {2013b}]{clark13b} Clark J., Ritchie B. \& Negueruela I., 2013b, A\&A, 560, A11

\bibitem[\protect\citeauthoryear{Clark {et al.}} {2014}]{clark14} Clark J., Ritchie B., Najarro F., Langer N. \& Negueruela I., 2014, A\&A, 565, A90

\bibitem[\protect\citeauthoryear{Danchi {et al.}} {2001}]{danchi01} Danchi W., Tuthill P. \& Monnier J., 2001, ApJ, 562, 440 

\bibitem[\protect\citeauthoryear{de Mink {et al.}} {2014}]{demink14} de Mink S., Sana H., Langer N., Izzard R. \& Schneider F., 2014, ApJ, 782, 7

\bibitem[\protect\citeauthoryear{Dougherty {et al.}} {2010}]{dough10} Dougherty S., Clark J., Negueruela I., Johnson T. \& Chapman J., 2010, A\&A, 511, A58

\bibitem[\protect\citeauthoryear{Fullerton, Massa \& Prinja {}} {2006}]{fullerton06} Fullerton A. Massa D. \& Prinja R. 2006, ApJ, 637, 1025

\bibitem[\protect\citeauthoryear{Gehrz {et al.}} {1995}]{gehrz95} Gehrz R., et al., 1995, ApJ, 439, 417

\bibitem[\protect\citeauthoryear{Gvaramadze \& Menten} {2012}]{gvar12} Gvaramadze V. \& Menten K., 2012 A\&A, 541, A7 

\bibitem[\protect\citeauthoryear{Humphreys \& Davidson} {1979}]{humphrey79} Humphreys R. M. \& Davidson K., 1979, ApJ, 232, 409



\bibitem[\protect\citeauthoryear{Kastner {et al.}} {2010}]{kastner10}Kastner J., Buchanan C., Sahai R., Forrest W. \& Sargent B., 2010, AJ, 139, 1993

\bibitem[\protect\citeauthoryear{Mart\'{i}n-Pintado {et al.}} {1989}]{martin89} Mart\'{i}n-Pintado J., Bachiller R., Thum C. \& Walmsley M., 1989, A\&A, 215, L13


\bibitem[\protect\citeauthoryear{Neguerela {et al.}} {2010}]{neg10} Negueruela I., Clark J. \& Ritchie B., 2010, A\&A, 516, A48

\bibitem[\protect\citeauthoryear{Petrovic {et al.}} {2005}]{petrovic05} Petrovic J., Langer N. \& van der Hucht K., 2005, A\&A, 435, 1013

\bibitem[\protect\citeauthoryear{Planesas {et al.}} {1992}]{planesas92} Planesas P., Martín-Pintado J. \& Serabyn E. 1992, ApJ, 386, L23

\bibitem[\protect\citeauthoryear{Podsiadlowski {et al.}} {2006}]{pods06} Podsiadlowski Ph., Morris T. S. \& Ivanova N., 2006, ASPC, 355, 259

\bibitem[\protect\citeauthoryear{Puls {et al.}} {2006}]{puls06} Puls J., Markova, N., Scuderi S., Stanghellini C., Taranova O., Burnley A. \& Howarth I., 2006, A\&A, 454, 625

\bibitem[\protect\citeauthoryear{Rau \& Cornwell {}} {2011}]{rau11} Rau U. \& Cornwell T. J., 2011, A\&A, 532, 71

\bibitem[\protect\citeauthoryear{Ritchie {et al.}} {2010}]{ritchie10} Ritchie B., Clark J., Negueruela I. \& Langer N., 2010, A\&A, 520, A48 

\bibitem[\protect\citeauthoryear{Rygl {et al.}} {2012}]{rygl12} Rygl K. et al., 2012, A\&A, 539, A79

\bibitem[\protect\citeauthoryear{Smith {et al.}} {2011}]{smith11} Smith N., Gehrz R. , Campbell R., Kassis M., Le Mignant D., Kuluhiwa K. \& Filippenko A., 2011, MNRAS, 418, 1959

\bibitem[\protect\citeauthoryear{Tafoya {et al.}} {2004}]{tafoya04} Tafoya D., G\'{o}mez Y. \& Rodr\'{i}guez L. F., 2004, ApJ, 610, 827 

\bibitem[\protect\citeauthoryear{Weintroub {et al.}} {2008}]{weintroub08} Weintroub J., Moran J. M., Wilner D. J., Young K., Rao R. \& Shinnaga H., 2008, ApJ, 677 1140 

\bibitem[\protect\citeauthoryear{White \& Becker {}} {1985}]{white85} White R. L. \& Becker R. H., 1985 ApJ 297 677 

\bibitem[\protect\citeauthoryear{Zickgraf {et al.}} {1985}]{zickgraf85} Zickgraf F.-J., Wolf B., Stahl O., Leitherer C. \& Klare G., 1985, A\&A, 143, 421

\end{thebibliography}
\end{document}